\documentclass[apjl]{emulateapj}

\usepackage{apjfonts}
\usepackage{mathptmx}

\usepackage{srcltx}
\usepackage{graphicx}
\usepackage{subfigure}

\begin{document}

\shorttitle{Alignment of Satellite Galaxies}
\shortauthors{Dong et al.}

\title{The Distribution of Satellites Around
  Central Galaxies in a Cosmological Hydrodynamical Simulation}

\author{X.C. Dong\altaffilmark{1,2}, W.P. Lin\altaffilmark{1},
  X. Kang\altaffilmark{3}, Yang Ocean Wang\altaffilmark{1,2}, 
Aaron A. Dutton\altaffilmark{4}, Andrea V. Macci\`o\altaffilmark{4}}

\altaffiltext{1}{Key Laboratory for Research in Galaxies and Cosmology,
Shanghai Astronomical Observatory, Chinese Academy of Science, 80 Nandan Road, 
Shanghai 200030, China}

\altaffiltext{2}{Graduate School, University of the Chinese Academy of
  Science, 19A, Yuquan Road, Beijing 100049, China}

\altaffiltext{3}{The Partner Group of MPI for Astronomy, Purple
  Mountain Observatory, 2 West Beijing Road, Nanjing 210008, China}

\altaffiltext{4}{Max Planck Institut f\"ur Astronomie, K\"onigstuhl 17, D-69117
Heidelberg, Germany}

\email{linwp@shao.ac.cn, kangxi@pmo.ac.cn}

\begin{abstract}

Observations have shown that the spatial distribution of satellite galaxies is not random, but rather is aligned with
the major axes of central galaxies (CGs). The strength of the alignment is dependent on the properties of both the
satellites and centrals. Theoretical studies using dissipationless $N$-body simulations are limited by their inability to
directly predict the shape of CGs. Using hydrodynamical simulations including gas cooling, star formation, and
feedback, we carry out a study of galaxy alignment and its dependence on the galaxy properties predicted directly
from the simulations.We found that the observed alignment signal is well produced, as is the color dependence: red
satellites and red centrals both show stronger alignments than their blue counterparts. The reason for the stronger
alignment of red satellites is that most of them stay in the inner region of the dark matter halo where the shape of
the CG better traces the dark matter distribution. The dependence of alignment on the color of CGs arises from
the halo mass dependence, since the alignment between the shape of the central stellar component and the inner
halo increases with halo mass. We also find that the alignment of satellites is most strongly dependent on their
metallicity, suggesting that the metallicity of satellites, rather than color, is a better tracer of galaxy alignment on
small scales. This could be tested in future observational studies.

\end{abstract}
\keywords{dark matter --- galaxies: formation --- galaxies: halo --- methods: numerical ---methods: statistical}

\section{Introduction}
\label{sec:intro}

Observations over the past decades have clearly shown that satellite
galaxies (SGs) are not randomly distributed, but rather are anisotropically
distributed around centrals.  This characteristic is observed
from our Milky Way Galaxy \cite[e.g.,][]{hol69, kro05}, the neighboring M31
\citep{gre99, koch06, iba13}, to large samples of local galaxies, and even in the Virgo cluster\citep{lee14}. 
In particular, both the results of 2dFGRS and Sloan Digital
Sky Survey (SDSS) have shown that
satellites are preferentially distributed along the major axes of
centrals. This phenomenon is known as galaxy alignment
\cite[e.g.,][]{sal04, bra05, Yang06}.  The alignment strength also
depends on the properties of both the satellites and centrals such that
red satellites show stronger alignment with centrals than blue
satellites, and red centrals have stronger alignment with their
satellites than blue centrals.  Such an alignment is also observed for
high-redshift galaxies \citep{Wang10,Nie12}.

Several groups have used theoretical arguments and 
numerical work to explain the origin of this alignment.
Studies of the Milky Way and
M31 have focused on the nature of the thin disk-like configuration of the
satellites \cite[e.g.,][]{Kang05, lib05, lib07, font11, metz07,
  lib12}, but debate exists concerning the rareness of such a distribution in
the CDM model due to the limited number of satellites and host systems
observed \cite[e.g.,][]{bahl14, iba14}.  More converged
conclusions are reached in the studies of galaxy alignment found in
the SDSS survey \cite[e.g.,][]{Kang07, fal07, bai08, agu10,
  Zhang13, Wang14a}.  The preferential distribution of satellites along
the major axes of centrals is found to be common in the cold dark matter (CDM)
model and arises from the non-spherical nature of dark halos
\cite[e.g.,][]{Jing2002}.

However, most studies so far have used $N$-body simulations where one
has to make assumptions about how the shapes of central galaxies (CGs) are
related to the shapes of their host halos. In most cases,
the predicted alignment signal is larger than observed if the CG shape follows the overall shape of the dark halo.
Furthermore, to explain the dependence of alignment on the galaxy
properties, one needs to adopt different assumptions for the shape of
centrals with blue and red color \citep{Kang07, agu10}.  To
directly predict the alignment signal, one should use simulations
which can self-consistently model the shapes of the centrals and
the distributions of the satellites (rather than the sub-halos).

\begin{figure*}[t]
 \epsscale{0.63}
 \plotone{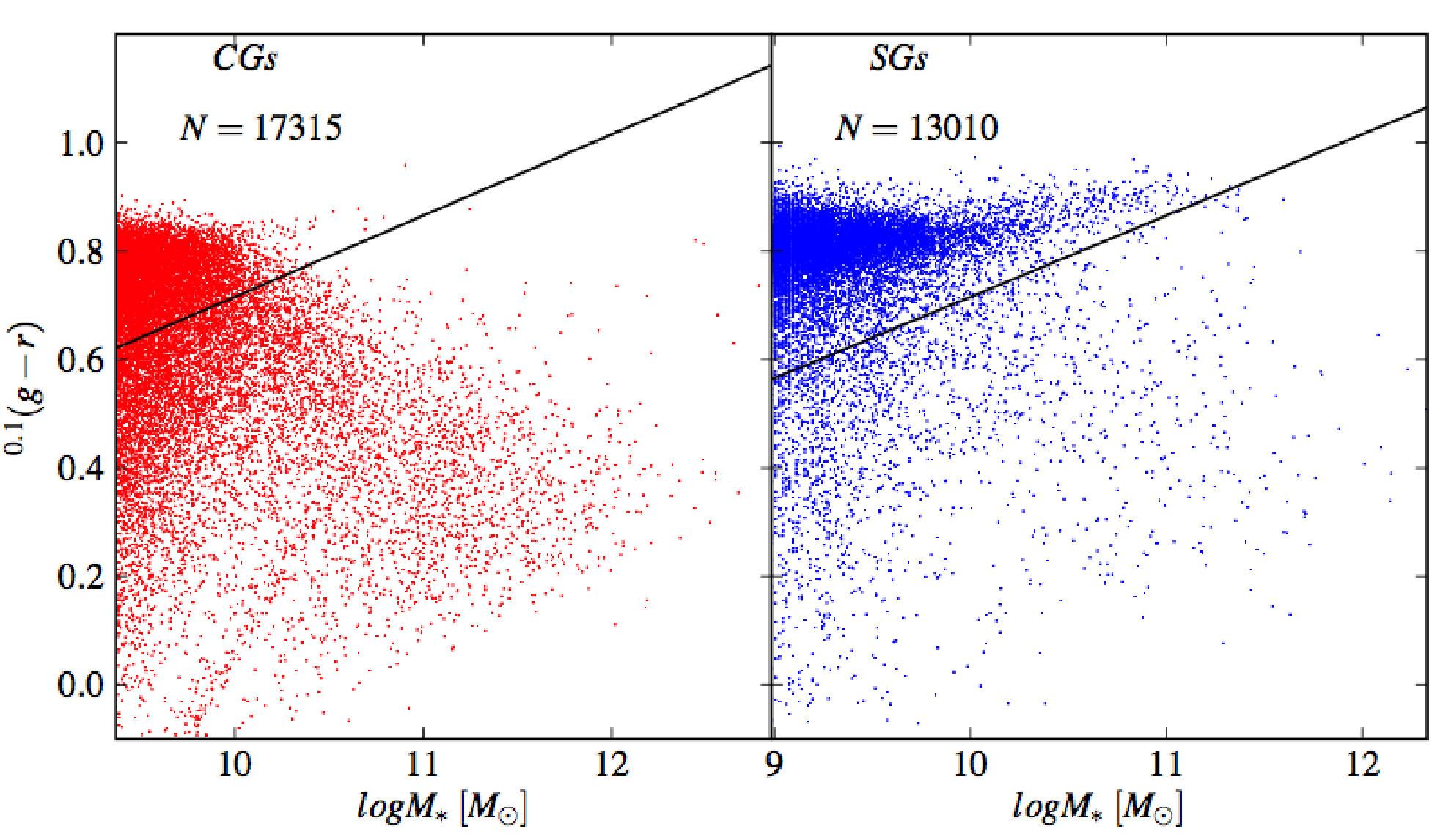}
 \caption{Relations between the color and stellar mass for our simulated
   galaxies: central galaxies (CG, left) and satellite galaxies (SG,
   right). The solid line is used to divide the populations into
   ``red'' and ``blue''.}
 \label{Color-plot}
\end{figure*}

\cite{Wang14a} used $N$-body simulations to study the dependence of 
alignment strength on halo assembly history.  
Due to the limited resolution and the lack of
gas physics, most sub-halos suffered from strong tidal stripping and
rarely survived in the central regions of the host halos. 
In this paper, we will study galaxy alignment using a
smoothed-particle-hydrodynamics (SPH) simulation which includes gas
cooling, metal recycling, star formation, and supernova feedback. As
the stellar component is included, the shape of the CG is
properly determined by the physics of gravity. The color of galaxies
can be easily obtained from the star formation history and metal
evolution from the simulation, using a stellar population synthesis
model and an assumed stellar initial mass function.  Compared to
$N$-body simulations, SPH simulations typically resolve more SGs
in the halo center, enabling the study of galaxy alignment on
smaller scales. This is because the gas cooling and subsequent star
formation results in a more compact mass distribution (than dark
matter only simulations) which is better able to survive the strong
tidal force in the host halo \citep{Sch11}.

There are a few studies which used SPH simulations to explore the galaxy alignment. For example, Libeskind et al. (2007) used high-resolution simulations of nine Milky Way like halos to study the spatial distribution of satellites, but mainly focused on their spatial configuration (polar disk). 
Deason et al. (2011) have investigated the alignment between the satellites' positions and CG using zoomed-in simulations, however, they focused on low-mass halos and did not  explore the dependence of alignment on satellite properties or compare with the data. 
Recently, Tenneti et al. (2014) utilized a high-resolution simulations with active galactic nucleus (AGN) feedback, but they only studied the shape correlation between dark matter halos and the stellar component. 
In this study, we will focus on the galaxy alignment with a dependence on the galaxy properties directly from our simulation, and also compare the model results with observational data (Yang et al. 2006) to understand what is the origin of the observed dependence.

\section{The simulation}

The cosmological simulation used in this paper was run using the
non-public version (including gas physics) of the massive parallel  code
Gadget-2 \citep{Springel2005}.  
It is evolved from redshift $z=120$ to the present epoch in
a cubic box of $100 h^{-1}{\rm Mpc}$ with $512^3$ of dark matter and gas
particles, assuming a flat $\Lambda{\rm CDM}$ ``concordance''
cosmology with $\Omega_m=0.268$, $\Omega_{\Lambda}=0.732$,
$\sigma_8=0.85$, and $h=0.71$.  A Plummer softening length of
$4.5 {\rm kpc}$ was adopted.  Each dark matter particle has a mass of
$4.62\times10^8 h^{-1}{M_{\odot}}$. The initial mass of gas
particles is $9.20\times10^7 h^{-1}{ M_{\odot}}$ and one gas
particle can turn into two star particles later on.  The simulation
includes the processes of radiative cooling, star formation,
supernova feedback, and outflows by galactic winds, as well as a
sub-resolution multiphase model for the interstellar medium.  The
readers are referred to \cite{Springel2003} for more details about
the treatment of gas physics.

Dark matter halos were found using the standard friends-of-friends
(FoF) algorithm with a linking length of 0.2 times the mean particle
separation, while the `galaxies' were defined as the stellar FoF
groups with a linking length of $4.88 h^{-1} {\rm kpc}$
\citep[c.f.][]{Jiang2008}.  The CG was identified as the
most massive galaxy (in stars) in the FoF halo; we call the other galaxies in the halo
SGs. To ensure the accurate determination of the shape of
centrals and to exclude spurious satellites, only those centrals and
satellites with more than $50$ and $20$ star particles, respectively, are
included in our sample.

For each star particle, we record its formation time and the
metallicity ([Fe/H]) of the cold gas at that time.  The luminosity of each star
particle at $z=0$ is calculated using the stellar population synthesis
model of \cite{BC03} assuming the initial mass function from
\cite{cha03}.  The sum of all the stellar particles gives the total
luminosity of each galaxy.  In our simulation, we calculate the magnitudes
in the SDSS $g$ and $r$ bands.

Figure 1 shows the predicted color-stellar mass diagram from our
simulation. The left and right panels are for the centrals and satellites,
respectively.  Here, $^{0.1}(g-r)$ is the galaxy color using filters
redshifted to $z=0.1$ and computed from $(g-r)$ using the method
described in \cite{Yang06}. Here we do not apply any dust
extinction to the model galaxies.  The solid line is the fit to the
SDSS data by \cite{vdB08} which divides galaxies into red and
blue. Compared to observations, the massive centrals in the
simulation are too blue.  This is a well-known problem of galaxy
formation models; without additional feedback, the gas cooling and
star formation in massive halos is too efficient, resulting in
massive centrals that are too blue
\cite[e.g.,][]{saro06}. With the inclusion of AGN feedback, the galaxy
stellar masses and colors in massive halos can be reproduced
\citep{Vog13}.

The right panel shows that the majority of satellites in our
simulation are red.  It is known that the main mechanisms for halting
star formation in satellites are ram-pressure stripping and the
intrinsic starvation by star formation using up the available gas,
both of which are included in our simulations.

Yang et al.  (2006) studied the dependence of galaxy alignment on
color. They simply adopted $^{0.1}(g-r)=0.83$ to divide galaxies
into red and blue. In fact, it is more reasonable that the division
between red and blue galaxies is stellar mass dependent (e.g., Baldry
et al. 2006; van den Bosch et al.2008; Kang \& van den Bosch 2008). 
Thus, in our work, we divide the galaxies into red and blue using the fitting line of van den Bosch et
al. (2008; the solid line in Figure 1).

\begin{figure*}[t]
  \epsscale{0.62}
  \plotone{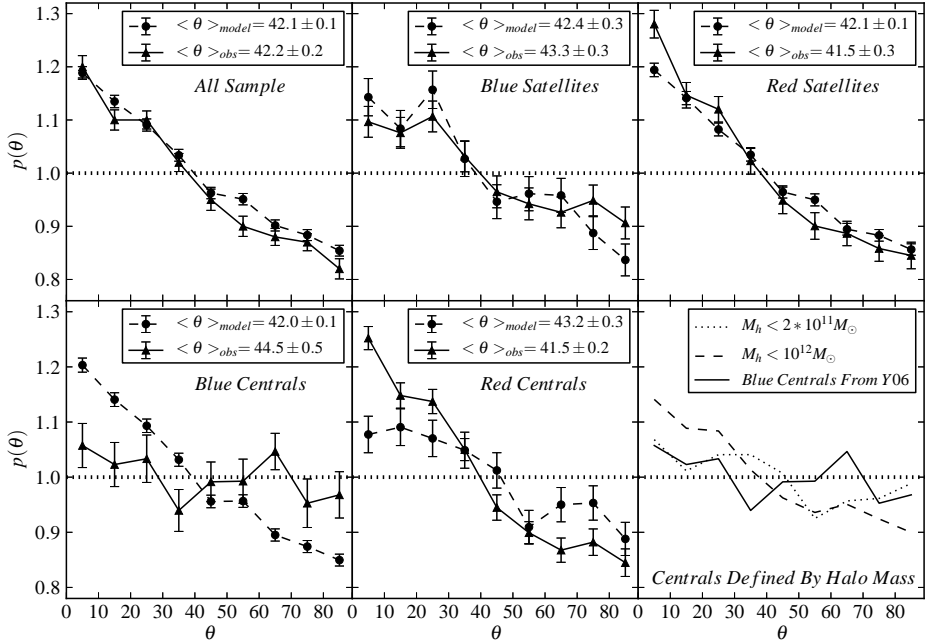}
  \caption{Predicted galaxy alignment (circles and dashed lines)
    and comparisons with observational results (triangles and solid
    lines). The upper left panel is for all galaxies, and the lower right
    panel is for central galaxies with different host halo masses. Other
    panels are for satellites and centrals with red/blue colors. The
    average alignment angle of the observed and model galaxies are labelled
    in each panel.}
    \label{compare_yang2006}
\end{figure*}

\section{Results}

To compare with the data more consistently, we measure the alignment
signal in projection. We project the model galaxies along one axis of
the simulation box and obtain the reduced inertia tensor of CGs 
from their stellar particles with $I_{ij} \equiv
\Sigma_{n}x_{i,n}x_{j,n}/r^2_{ij,n}$ where $x_{i,n}$ is the position
of the $n$ th stellar particle in the halo center coordinates and
$r_{ij,n}=\sqrt{x^2_{i,n}+x^2_{j,n}}$.  The eigenvectors of $I_{ij}$
define the orientation of the galaxy, and with this form of the
inertia tensor the major axis corresponds to the vector with the
large eigenvalue. The angle $\theta$ ($0^{\circ} \leq \theta \leq
90^{\circ}$) is defined as the angle on the projection plane between
the off-center vector of the satellite position and the major axis of the
CG.  The distribution probability of $\theta$ is defined
as $P(\theta) = \frac{N(\theta)} {<N_{R}(\theta)>}$ where $N(\theta)$
and $<N_{R}(\theta)>$ are the number of central-satellite pairs in the
simulation sample and random samples.  An alignment is found if
$<P(\theta)\theta>$ is less than $45^{\circ}$.  The bootstrapping
method is used to derive $<N_{R}(\theta)>$ by randomly selecting 100
sub-samples  from the overall data.

The predicted alignment and comparisons with data are shown in
Figure \ref{compare_yang2006}. The triangles with solid lines are the
observational results of Yang et al. (2006) and model results are
plotted with circles connected by dashed lines. The upper left panel
shows that the predicted alignment effect for all galaxies agrees well
with the data. For satellites, the model predictions roughly
agree with the data, but slightly overpredict the alignment for blue
satellites and underpredict the alignment for red satellites.
However, the predicted alignment for centrals is inconsistent
with the data. The lower left and middle panels show that the blue
centrals have stronger alignment than the red ones, contrary to the
observed dependence on color.
 
As seen in Figure 1 and the discussion above, there are too many massive
blue galaxies in our simulation due to absence of an effective
quenching mechanism in massive haloes (such as AGN feedback). It is
commonly believed that these massive galaxies should be red and exist
in massive halos.  Both theoretical and observational works suggest
that there should be a critical halo mass below which most of the
centrals are blue, and above which most are red \citep{Yang08,  ker05}. 
However, this critical halo mass is not well determined, ranging from $2\times 10^{11}M_{\odot}$ 
\cite[e.g.,][]{ker05} to $10^{12}M_{\odot}$ \cite[e.g.,][]{cat08}.  
To simply illustrate this effect, in the lower right panel,
we show the predicted alignment for blue centrals that have a halo mass less than
$M_{c}$, where $M_{c}=2\times 10^{11}M_{\odot}$ (dotted line) and
$M_{c}=10^{12}M_{\odot}$ (dashed line). It is found that the
predicted alignment for blue centrals is close to the data with $M_{c}=2\times
10^{11}M_{\odot}$, and it increases with increasing critical halo mass. 
To avoid overlapping, we do not plot the lines for red centrals whose halo mass are larger than $M_c$. 
The results are close to the observational one  and the lines for the two
critical mass are almost identical, since  massive halos have more satellites than low-mass halos and 
thus dominate the signal.

The exercise presented in the lower right panel of Figure 2 suggests that
the observed alignment dependence on CG color is mainly
determined by the host halo mass.  Later, we will  see that the halo
mass dependence is rooted in the correlation between the shapes of the
CG and the host halo.

\begin{figure*}[t]
  \epsscale{.62}
  \plotone{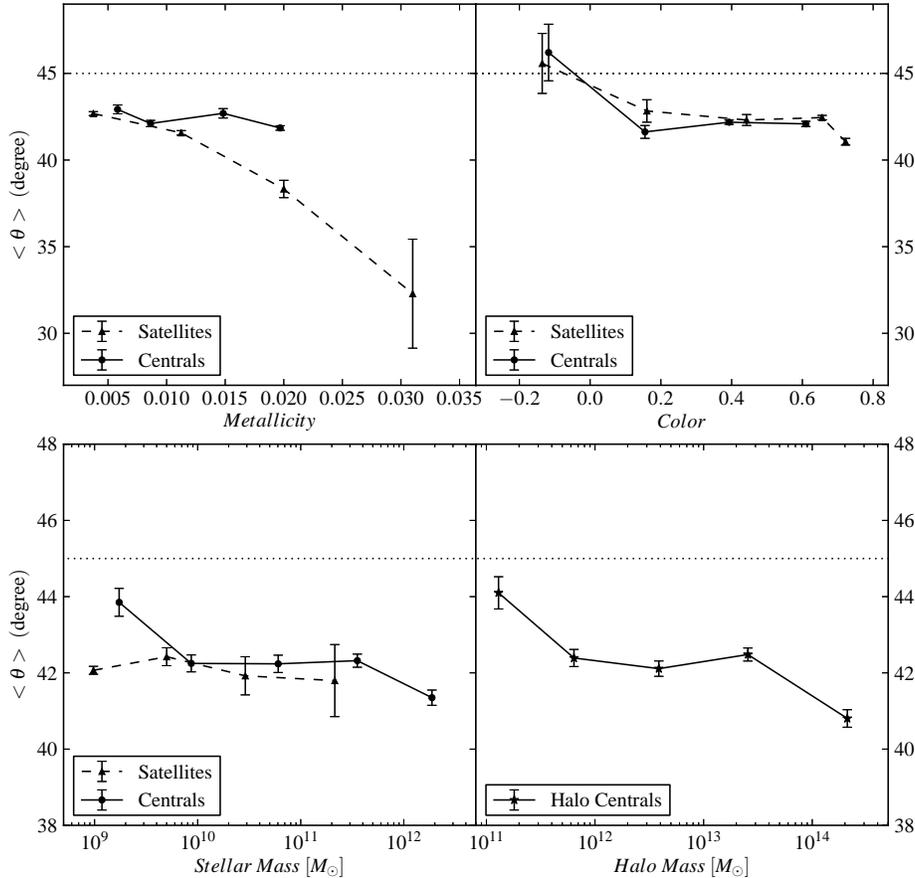}
  \caption{Dependence of alignment strength (2D) on the properties
    of simulated galaxies and dark halos. See the text for more
    details. }
    \label{dependence}
\end{figure*}

In Figure \ref{dependence}, we further show the dependence of alignment
on galaxy properties (metallicity, color, stellar mass, and halo mass)
from the simulation. The solid and dashed lines are for the centrals and satellites, respectively. 
The upper panels show that the satellite alignment depends on the metallicity and color, with a stronger
dependence on metallicity for which metal-rich satellites have very strong
alignment.  The lower left panel shows that the dependence on stellar
mass is  weak and fainter satellites have slightly weaker
alignment, in broad agreement with the finding in Y06.  Note that the
error-bar for the point at the greatest mass bin is large due to
the small number statistics. 
The halo mass dependence in the lower right panel is close to the dependence from
the data -- there is stronger alignment in more massive halos, which
is also consistent with the dependence on stellar mass for centrals in the left panel.

\begin{figure*}[htbp]
  \epsscale{.62}
  \plotone{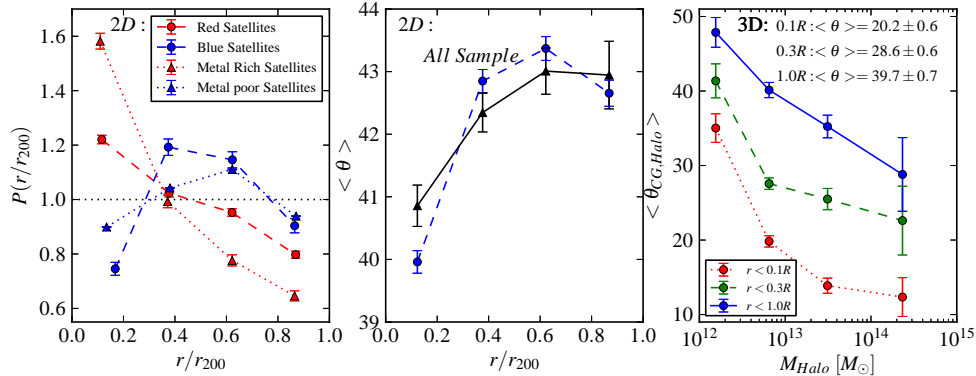}
  \caption{Left panel:  radial distribution of red and blue,  metal-rich (top 30\% by order ranking), and 
     metal poor (bottom 30\% by order ranking) SGs within dark matter halos. 
     Middle panel: dependence of average  alignment angle on radius from the halo center.  
     Right panel: distribution of mis-alignment angles between the major axes of CGs
    and those of dark halos measured within radii of $0.1$, $0.3$, and $1.0 R_{200}$.}
    \label{alignment_BCG_halo}
\end{figure*}

In addition to the dependence of galaxy alignment on color,
observations have shown that the alignment angle is a strong
function of radial distance to the centrals \citep{bra05, Yang06}. 
In the left panel of Figure \ref{alignment_BCG_halo}, we show the radial
spatial distribution of satellites in the dark halo, together
with the dependence on color and metallicty.  It is found that both
metal-rich (top 30\% by order ranking) and red satellites are distributed predominately in the
central halo.  
In general, this distribution agrees with the observational facts
that galaxy properties, such as color, metallicity, or morphology,
depend strongly on environment/local density (e.g., Wang et al. 2014b and reference therein). 
Such trends are expected since metal recycling and star formation quenching are more
efficient in the inner halo \citep{Wein10}.  
The middle panel in Figure \ref{alignment_BCG_halo} shows the alignment of satellites as
a function of the radius.  The observational results of Y06 are shown with
triangles and a solid line, which show that satellites residing
in the inner halo have stronger alignment with the CG than their
counterparts residing in the outer halo. Our simulation results (blue
circles and dashed line) are in good agreement with these
observations.

To understand the origin of satellite alignment with respect to the
CG, theoretical work using $N$-body simulations often assumed
that the shape of the CG follows the shape of the overall dark
matter halo. However, this assumption leads to a stronger alignment
than observed \cite[e.g.,][]{Kang07}.  To decrease the predicted
signal, one has to introduce some degree of mis-alignment between the
CG and that of the dark matter halo \citep{Kang07, agu10}.
A more physical solution is that the CG follows the shape of
dark matter in the central region. This model is better at reproducing
the alignment \citep{fal09, Wang14a}, however, the dependence on galaxy
color has not been reproduced.

As our SPH simulation includes the stellar component, we can directly
predict the shape of the CG and  test the assumption. 
The right panel of Figure \ref{alignment_BCG_halo} shows the
average alignment angle between the major axis of the CG
and the shape of the dark matter halo as a function of the halo mass.
Results are shown for halo shapes calculated within three radii: for
the ``whole halo'' inside of $R_{200}$ (blue solid), the
``intermediate halo'' inside of $0.3R_{200}$ (green dashed), and the
``inner halo'' inside of $0.1 R_{200}$ (red dotted).  Here, $R_{200}$
is the spherically averaged virial radius.  It is found that the shape
of the CG better traces  that of the inner halo, and this
alignment increases with halo mass.  The mean mis-alignment angle
varies from $\sim 35 \- 10 \deg$ for halos with a mass of $\sim
10^{12}M_{\odot} \- 10^{14}M_{\odot}$. Similar results are also found
in recent work \cite[e.g,][]{tenneti14}.

The results in Figure \ref{alignment_BCG_halo} clearly explain the
observed galaxy alignment dependence on galaxy properties.  The most
metal rich and reddest satellites are distributed in the inner region of
the host halo, and their spatial distribution should
closely follow that of dark matter.  On the other hand, the stellar
component of centrals is also greatly shaped by the
gravitational force of the dark matter in the inner halo.  The
combination of these two effects leads to a better alignment for the metal-rich/red satellites than
their metal-poor/blue counterparts.  As to the dependence on color of
centrals, this is related to the halo mass of the centrals -- bluer
centrals most likely reside in relatively lower-mass halos where the
alignment between the central stellar component and the inner halo shape
becomes weaker.

\section{Conclusion and Discussion}

In this Letter, we carry out a study of galaxy alignment using
a cosmological simulation including gas cooling, star
formation, and supernova feedback, which enables a direct prediction
for the shape of CGs and the galaxy properties.  
We find that the predicted alignment between the CG and the distribution of satellites 
agrees with the observations. Furthermore, with a simple
assumption about the halo mass of blue and red centrals, the
dependence on color for both centrals and satellites is also reproduced.
We also identify that the strongest dependence of the alignment is with
metallicity of satellites, which should be testable using future data.

The main source of galaxy alignment is the non-spherical nature
of CDM halos, as shown by many previous studies
\cite[e.g.,][]{agu06a,agu06b, Kang07}.  However, the predicted strength
of the alignment is too strong if the shape of the CG
follows the overall shape of the dark matter halo. From our study, we find
 that the shape of the CG better follows the halo in
the inner region, and the average mis-alignment is about $20 \deg$
(see Figure \ref{alignment_BCG_halo}), similar to the expected or
inferred values in previous studies \cite[e.g.,][]{Wang08,fal09}.  As
the most red/metal-rich satellites stay in the inner halo, they
naturally follow the shape of the dark matter halo in that region.  This
leads to a strong alignment between red satellites with
centrals. Furthermore, as the alignment between the CG and inner
halo increases with halo mass and red centrals predominately
populate massive halos, it explained the observed fact that red
central shows stronger alignment with satellites than blue centrals.
Although the prediction for the alignment of blue centrals using our
simulation fails because of the too blue colors of the most massive central galaxies,
the exercises for the alignment dependence on halo mass have given
hints that simulations with AGN feedback \cite[e.g.,][]{Vog13,tenneti14} should
be helpful to solve this problem.

The non-spherical nature of dark matter halos is one the most
prominent features of structure formation in a CDM
universe, as the mass accretion and mergers predominately occur along
the cosmic web or the filament \cite[e.g.,][]{Wang05}.  It also
naturally produces the galaxy alignment on very large scales up to
$\sim 70 h^{-1} {\rm Mpc}$ \citep{Li13}.  Accurate predictions for galaxy
alignment on large scales is crucial to cosmological applications,
such as estimating the systematic error used in weak lensing
measurements.  With the proper modeling of galaxy shapes from
hydrodynamical simulation, we will be able to make predictions for
galaxy alignment on large scales in a forthcoming paper.

\acknowledgments
The authors thank the anonymous referee for useful suggestions. 
W.P.L. and X.K. acknowledge supports by the NSFC projects (No. 11121062,
11233005, U1331201, 11333008) and the ``Strategic Priority Research
Program the Emergence of Cosmological Structures'' of the Chinese
Academy of Sciences (grant No. XDB09010000).  X.K. is also supported by
the National basic research program of China (2013CB834900) and 
the foundation for Distinguished Young Scholars of Jiangsu Province (No. BK20140050) . 
The simulations were run in the Shanghai Supercomputer Center and the data
analysis was performed on the supercomputing platform of Shanghai
Astronomical Observatory.
X.K., A.A.D. and A.V.M., acknowledge support from
the MPG-CAS through the partnership program between
the MPIA group lead by A. Macci\`o and the PMO group lead by X. Kang.

\end{document}